# General Solutions of the Abel Differential Equations

Ji-Xiang Zhao
College of Information Engineering, China Jiliang University, Hangzhou, 310018, China
zhaojixiang@cjlu.edu.cn, ORCID: 0000-0003-2426-8741

**Abstract:** The Abel differential equations play a significant role in various fields of mathematics and applied sciences and are classified into two types: the first kind and the second kind. A novel derivative condition for the general solution of first-kind Abel equation is introduced. Based on this condition, the general solutions to the first-kind Abel equation with a zero free term are obtained, which in turn enables the derivation of the general solutions to the second-kind Abel equation, and meanwhile, a pair of entangled functions is discovered. These results can be extended to the Liénard equation.

**Keywords**: analytical general solutions; the Abel differential equation; entangled functions

**MSC code**: 34A34, 34A05, 34A25

## 1. Introduction

The Abel differential equation includes the first kind

$$y' = f_3(x)y^3 + f_2(x)y^2 + f_1(x)y + f_0(x)\,,\ y = y(x)\,,\ f_3 \neq 0 \tag{1}$$

and the second kind

$$[g_1(x)Y + g_0(x)]Y' = F_2(x)Y^2 + F_1(x)Y + F_0(x)\,,\ Y = Y(x) \tag{2}$$

considerable research effort have been attracted because they are playing significant roles in two aspects: one is that it is deeply related to classical mathematics, such as two classical problems on plane polynomial vector fields: Hilbert's 16th problem and Poincare's center-focus problem [1, 2]; the other is in the field of applied science [3]. However, general solution of the Abel equation has not been obtained except in some special cases so far [4-6].

In this paper, we will give the analytical solution to the Abel equation.

## 2. Preliminary

We give an auxiliary result, which will be used to prove the main result.

**Lemma** For general form of Abel equation of the first kind

$$y' = a(x)\,y^3 + b(x)\,y^2 + c(c)\,y + d(x), \quad y = y(x) \tag{3}$$

there exit a differential function $w(x)$ and constant $k$ , such that

$$d = c\left(\frac{w}{3}\right) + kb\left(\frac{w}{3}\right)^2 - (3k+2)\,a\left(\frac{w}{3}\right)^3 - \left(\frac{w}{3}\right)' \tag{4}$$

and

$$\left(\frac{b}{a} - w\right)' = -\frac{2(k+1)b}{9\lambda}\left(\frac{b}{a} - w\right)^2 + \left[c + \left(\frac{k+1}{3\lambda} - 1\right)\frac{b^2}{3a}\right]\left(\frac{b}{a} - w\right) \tag{5}$$

then, general solution of equation (3) can be exactly given as

$$y(k,x) = -\frac{\sqrt[3]{3k+2}}{3}\,\frac{e^{\int\left[c + \left(\frac{k+1}{3\lambda} - 1\right)\frac{b^2}{3a}\right]dx}}{c_0 + \dfrac{2(k+1)}{9\lambda}\displaystyle\int b e^{\int\left[c + \left(\frac{k+1}{3\lambda} - 1\right)\frac{b^2}{3a}\right]dx}\,dx} - \frac{b}{3a} \tag{6}$$

where $\lambda = -\dfrac{1 + \sqrt[3]{3k+2}}{3}$ and $c_0$ is an integration constant.

**Proof**: Substituting formula (4) into equation (3) yields

$$\left(y + \frac{w}{3}\right)' = a\left(y + \frac{w}{3}\right)^3 + a\left(\frac{b}{a} - w\right)\left(y + \frac{w}{3}\right)^2 + \left[\left(c - \frac{b^2}{3a}\right) + \frac{a}{3}\left(\frac{b}{a} - w\right)^2\right]\left(y + \frac{w}{3}\right)$$
$$+ \frac{k+1}{9}\,a\left[\left(\frac{b}{a} - w\right)^3 - \frac{2b}{a}\left(\frac{b}{a} - w\right)^2 + \frac{b^2}{a^2}\left(\frac{b}{a} - w\right)\right]$$

let

$$y + \frac{w}{3} = \lambda\left(\frac{b}{a} - w\right) \tag{7}$$

where $\lambda(x) \neq 0$ is a function to be determined, and above equation is transformed to

$$\begin{aligned}\left[\lambda\left(\frac{b}{a} - w\right)\right]' &= a\left[\left(1 + \frac{1}{3\lambda}\right)^3 + (3k+2)\left(\frac{1}{3\lambda}\right)^3\right]\left[\lambda\left(\frac{b}{a} - w\right)\right]^3 \\ &\quad - \frac{2(k+1)b}{9\lambda^2}\left[\lambda\left(\frac{b}{a} - w\right)\right]^2 + \left[c + \left(\frac{k+1}{3\lambda} - 1\right)\frac{b^2}{3a}\right]\left[\lambda\left(\frac{b}{a} - w\right)\right]\end{aligned} \tag{8}$$

set $\left(1 + \dfrac{1}{3\lambda}\right)^3 + (3k+2)\left(\dfrac{1}{3\lambda}\right)^3 = 0$, that is to say $\lambda = -\dfrac{1 + \sqrt[3]{3k+2}}{3}$, and equation (8) is

reduced to the constraint equation (5), we get

$$\frac{b}{a}-w=\frac{e^{\int\left[c+\left(\frac{k+1}{3\lambda}-1\right)\frac{b^2}{3a}\right]dx}}{c_0+\frac{2(k+1)}{9\lambda}\int be^{\int\left[c+\left(\frac{k+1}{3\lambda}-1\right)\frac{b^2}{3a}\right]dx}dx} \tag{9}$$

by the transformation equation (7), the general solution of the Abel equation (3) can be obtained and given as formula (6).

## 3. Main Results

### 3.1 The Abel equation of the first kind

In this section, we will give the general solutions of the first-kind Abel equation (1) with free term $f_0(x)=0$ by means of two theorems.

**Theorem.1** The Abel equation (1) is integrable with the constraint condition

$$\left(\frac{f_2}{f_3}\right)'=\frac{f_0}{\lambda}+\lambda(\lambda+1)\frac{f_2^3}{f_3^2}+\frac{f_1f_2}{f_3} \tag{10}$$

and general solutions can be exactly obtained, where $\lambda$ is a constant.

**Proof**: The substitution $y(x)=u(x)z(x)+v(x)$ brings equation (1) into

$$z'=f_3u^2z^3+(3f_3v+f_2)uz^2+\left(3f_3v^2+2f_2v+f_1-\frac{u'}{u}\right)z+\frac{f_3v^3+f_2v^2+f_1v+f_0-v'}{u}$$

by the **Lemma**, if

$$\begin{aligned}\frac{f_3v^3+f_2v^2+f_1v+f_0-v'}{u}=&\left(3f_3v^2+2f_2v+f_1-\frac{u'}{u}\right)\frac{w}{3}\\&+k(3f_3v+f_2)u\left(\frac{w}{3}\right)^2-(3k+2)f_3u^2\left(\frac{w}{3}\right)^3-\left(\frac{w}{3}\right)'\end{aligned} \tag{11}$$

and

$$\begin{aligned}\left[\frac{1}{u}\left(3v+\frac{f_2}{f_3}-uw\right)\right]'=&-\frac{2(k+1)(3f_3v+f_2)u}{9\lambda}\left[\frac{1}{u}\left(3v+\frac{f_2}{f_3}-uw\right)\right]^2\\&+\left[3f_3v^2+2f_2v+f_1-\frac{u'}{u}+\left(\frac{k+1}{3\lambda}-1\right)\frac{(3f_3v+f_2)^2}{3f_3}\right]\left[\frac{1}{u}\left(3v+\frac{f_2}{f_3}-uw\right)\right]\end{aligned} \tag{12}$$

hold, we have

$$z(x)=-\frac{1}{u}\left\{\frac{\sqrt[3]{3k+2}}{3}\frac{e^{\int\theta dx}}{C_0+\frac{2(k+1)}{9\lambda}\int(3f_3v+f_2)e^{\int\theta dx}dx}+v+\frac{f_2}{3f_3}\right\} \tag{13}$$

where

$$\theta(x)=f_1-\frac{f_2^2}{3f_3}+\frac{k+1}{3\lambda}\frac{(3f_3v+f_2)^2}{3f_3} \tag{14}$$

we rewrite equations (11) and (12) as, respectively

$$\begin{aligned}\left(v-\frac{uw}{3}\right)'&=f_3\left(v-\frac{uw}{3}\right)^3+f_2\left(v-\frac{uw}{3}\right)^2+\left[f_1-3(k+1)f_3\left(\frac{uw}{3}\right)^2\right]\left(v-\frac{uw}{3}\right)\\&\quad-(k+1)f_2\left(\frac{uw}{3}\right)^2+f_0\end{aligned} \tag{15}$$

and

$$\begin{aligned}\left(3v+\frac{f_2}{f_3}-uw\right)'&=-\frac{2(k+1)(3f_3v+f_2)}{9\lambda}\left(3v+\frac{f_2}{f_3}-uw\right)^2\\&\quad+\left[3f_3v^2+2f_2v+f_1+\left(\frac{k+1}{3\lambda}-1\right)\frac{(3f_3v+f_2)^2}{3f_3}\right]\left(3v+\frac{f_2}{f_3}-uw\right)\end{aligned} \tag{16}$$

and suppose that $-(k+1)f_2\left(\frac{uw}{3}\right)^2+f_0=0$, i.e. $\frac{uw}{3}=\pm\sqrt{\frac{f_0}{(k+1)f_2}}$, from equation (15), we have

$$v=\frac{uw}{3}=\pm\sqrt{\frac{f_0}{(k+1)f_2}}\quad,(k+1\neq 0) \tag{17}$$

and equation (16) is reduced to the constraint equation (10). Thus, by the transformation $y=uz+v$, the general solutions of the Abel equation (1) can be given by

$$y(x)=-\left[\frac{\sqrt[3]{3k+2}}{3}\frac{e^{\int\theta dx}}{c_0+\frac{2(k+1)}{9\lambda}\int\left(f_2\pm 3f_3\sqrt{\frac{f_0}{(k+1)f_2}}\right)e^{\int\theta dx}dx}+\frac{f_2}{3f_3}\right] \tag{18}$$

where $\theta(x)=f_1-\dfrac{f_2^2}{3f_3}+\dfrac{k+1}{9\lambda f_3}\left(f_2\pm 3f_3\sqrt{\dfrac{f_0}{(k+1)f_2}}\right)^2$.

**Theorem.2** If $f_0(x)=0$, the Abel equation (1) is integrable and general solutions can be obtained exactly.

**Proof**: The substitution $y(x)=\phi(x)+\varphi(x)$ brings equation (1) into

$$\phi'=f_3\phi^3+(3f_3\varphi+f_2)\phi^2+\left(3f_3\varphi^2+2f_2\varphi+f_1\right)\phi+f_3\varphi^3+f_2\varphi^2+f_1\varphi+f_0-\varphi'$$

by the theorem 1, know that when

$$\begin{aligned}\left(3\varphi+\frac{f_2}{f_3}\right)'&=\frac{1}{\lambda}\left(f_3\varphi^3+f_2\varphi^2+f_1\varphi+f_0-\varphi'\right)+\lambda(\lambda+1)\frac{(3f_3\varphi+f_2)^3}{f_3^2}\\&\quad+\left(3f_3\varphi^2+2f_2\varphi+f_1\right)\left(3\varphi+\frac{f_2}{f_3}\right)\end{aligned}$$

which can be rewritten as

$$\begin{aligned}\left[(3\lambda+1)\varphi\right]'&=f_3\left[(3\lambda+1)\varphi\right]^3+(3\lambda+1)f_2\left[(3\lambda+1)\varphi\right]^2\\&\quad+\left[f_1+\lambda(3\lambda+2)\frac{f_2^2}{f_3}\right](3\lambda+1)\varphi\\&\quad+f_0+\lambda^2(\lambda+1)\frac{f_2^3}{f_3^2}+\lambda\left[\frac{f_1f_2}{f_3}-\left(\frac{f_2}{f_3}\right)'\right]\end{aligned}\tag{19}$$

holds, there is

$$\phi(x)=-\left[\frac{\dfrac{\sqrt[3]{3k+2}}{3}e^{\int\theta dx}}{c_0+\dfrac{2(k+1)}{9\lambda}\displaystyle\int\left(3f_3\varphi+f_2\pm 3f_3\sqrt{\dfrac{f_3\varphi^3+f_2\varphi^2+f_1\varphi+f_0-\varphi'}{(k+1)(3f_3\varphi+f_2)}}\right)e^{\int\theta dx}dx}+\varphi+\frac{f_2}{3f_3}\right]\tag{20}$$

where $\theta(x)=f_1-\dfrac{f_2^2}{3f_3}+\dfrac{k+1}{9\lambda f_3}\left(3f_3\varphi+f_2\pm 3f_3\sqrt{\dfrac{f_3\varphi^3+f_2\varphi^2+f_1\varphi+f_0-\varphi'}{(k+1)(3f_3\varphi+f_2)}}\right)^2$. Set $(3\lambda+1)\varphi=-\lambda\dfrac{f_2}{f_3}$, that is

$$\varphi(x) = -\frac{\lambda}{3\lambda+1}\frac{f_2}{f_3} \tag{21}$$

then equations (19) and (21) are reduced to, respectively

$$f_0(x) = 0 \quad \text{and}$$

$$\phi(x) = \frac{\frac{3\lambda+1}{3}e^{\int \theta dx}}{c_0 + \frac{2}{27}\frac{1-(3\lambda+1)^3}{\lambda}\int\left(\frac{f_2}{3\lambda+1} \pm 3 f_3\sqrt{A}\right)e^{\int \theta dx}dx} - \varphi - \frac{f_2}{3f_3} \tag{22}$$

where

$$A(x) \equiv \frac{3\lambda}{\left[1-(3\lambda+1)^3\right]f_2}\left[\frac{\lambda(2\lambda+1)}{(3\lambda+1)^2}\frac{f_2^3}{f_3^2} - \frac{f_1 f_2}{f_3} + \left(\frac{f_2}{f_3}\right)'\right] \tag{23}$$

and

$$\theta(x) = f_1 - \frac{f_2^2}{3f_3} + \frac{1-(3\lambda+1)^3}{27\lambda f_3}\left(\frac{f_2}{3\lambda+1} \pm 3 f_3\sqrt{A}\right)^2 \quad (\lambda \neq 0, -\frac{1}{3}) \tag{24}$$

Finally, by the transformation $y = \phi + \varphi$, the general solutions of equation (1) can be obtained and given by

$$y(\lambda, x) = \frac{(3\lambda+1)e^{\int \theta dx}}{c_0 + \frac{2}{9}\frac{1-(3\lambda+1)^3}{\lambda}\int\left(\frac{f_2}{3\lambda+1} \pm 3 f_3\sqrt{A}\right)e^{\int \theta dx}dx} - \frac{f_2}{3f_3} \tag{25}$$

**Remark**: This result can be used to construct a new superposition rule for Abel differential equation of the first kind, which involves only one particular solution.

### 3.2 The Abel equation of the second kind

In this section, based on the theorem 2 of the previous section, we will present the general solutions for the Abel equation of the second kind without any constraints.

**Theorem.3** The Abel's second kind equation (2) is integrable and general solutions can be obtained exactly.

**Proof**: The Abel equation (2) can be transformed into the first kind by the change $y = \left[g_1(x)Y + g_0(x)\right]^{-1}$ with coefficients

$$f_0(x)=0,\quad f_1(x)=-\frac{F_2+g_1'}{g_1}\ \text{and}$$

$$f_2(x)=g_0\left(\frac{g_1'}{g_1}+\frac{2F_2}{g_1}\right)-\left(g_0'+F_1\right),\quad f_3(x)=g_0\left(F_1-\frac{g_0}{g_1}F_2\right)-g_1F_0 \tag{26}$$

according to the theorem 2, the general solutions of equation (2) are given by

$$Y(\lambda,x)=\frac{1}{g_1}\left(\frac{1}{y}-g_0\right)$$

in which, $y(x)$ is given by formula (25) and $f_j(x)$ ( $j=1,2,3$ ) are replaced by formulas (26).

## 4. Concluding remark

The general solutions with free parameter $\lambda$ to the Abel equation are presented in this paper. In the process of solving, an entangled function pair, such as ($u$ and $w$) entangled as $\frac{uw}{3}=\pm\sqrt{A}$ is discovered.

It is worth to say, that these results can also be applied to solve the Liénard equation $\frac{d^2y}{dx^2}+f(y)\frac{dy}{dx}+g(y)=0$, which can be transformed to Abel differential equation $\frac{dw}{dy}=f(y)w^2+g(y)w^3$ by the transformation $\frac{1}{w(y)}=\frac{dy}{dx}$.

**Declaration of Interest Statement**

The author declares that there is no conflict of interests regarding the publication of this paper.

**Data availability**

No data were used for the research described in this paper.